# Severe Dirac Mass Gap Suppression in $Sb_2Te_3$-based Quantum Anomalous Hall Materials

Yi Xue Chong[1,2*], Xiaolong Liu[1,3*], Rahul Sharma[1,2], Andrey Kostin[1], Genda Gu[2], K. Fujita[2], J.C. Séamus Davis[1,4,5,#] and Peter O. Sprau[1,6]

1. LASSP, Department of Physics, Cornell University, Ithaca, NY 14853, USA
2. CMPMS Department, Brookhaven National Laboratory, Upton, NY 11973, USA
3. Kavli Institute at Cornell, Cornell University, Ithaca, NY 14853, USA
4. Department of Physics, University College Cork, Cork T12R5C, IE
5. Clarendon Laboratory, University of Oxford, Oxford, OX1 3PU, UK
6. Advanced Development Center, ASML, Wilton, CT 06897, USA.

* Y.X.C and X.L contributed equally to this work.
# Correspondence should be addressed to: jcseamusdavis@gmail.com

**Abstract**: Quantum anomalous Hall (QAH) effect appears in ferromagnetic topological insulators (FMTI) when a Dirac mass gap opens in the spectrum of the topological surface states (SS). Unaccountably, although the mean mass gap can exceed 28 meV (or ~320 K), the QAH effect is frequently only detectable at temperatures below 1 K. Using atomic-resolution Landau level spectroscopic imaging, we compare the electronic structure of the archetypal FMTI $Cr_{0.08}(Bi_{0.1}Sb_{0.9})_{1.92}Te_3$ to that of its non-magnetic parent $(Bi_{0.1}Sb_{0.9})_2Te_3$, to explore the cause. In $(Bi_{0.1}Sb_{0.9})_2Te_3$, we find spatially random variations of the Dirac energy. Statistically equivalent Dirac energy variations are detected in $Cr_{0.08}(Bi_{0.1}Sb_{0.9})_{1.92}Te_3$ with concurrent but uncorrelated Dirac mass gap disorder. These two classes of SS electronic disorder conspire to drastically suppress the minimum mass gap to below 100 μeV for nanoscale regions separated by <1 μm. This fundamentally limits the fully quantized anomalous Hall effect in $Sb_2Te_3$-based FMTI materials to very low temperatures.

**Main Text**:

The exemplary topological state of electronic matter is the integer quantum Hall (IQH) effect. This occurs when a high magnetic field generates Landau quantization in the states of a 2-dimensional electron gas, resulting in precisely quantized Hall conductance $\sigma_{xy} = Ce^2/h$ and transport resistivity $\rho_{xx} = 0$ ($C$ is the Chern number equaling the number of Landau levels (LL) below the Fermi energy $E_F$, $e$ is the electron charge and $h$ is Planck's constant). Even more striking is the quantum anomalous Hall (QAH) effect[1,2] which should support $\sigma_{xy} = e^2/h$ and $\rho_{xx} = 0$ in the absence of external magnetic fields and at any temperature where global ferromagnetism can be sustained. The advent of topological insulators (TI) with relativistic surface states (SS) made this feasible because, if these materials can be rendered ferromagnetic, the spontaneous magnetization gaps the SS Dirac spectrum, generating a QAH effect with its chiral edge states.[3]

Although the existence of the QAH in such ferromagnetic topological insulators (FMTI) has now been widely demonstrate,[4-15] its detailed phenomenology remains mysterious. This is because the fully quantized $\sigma_{xy} = \pm e^2/h$ is typically only detected at temperatures well below 1 K even though the average Dirac-mass gap $\Delta$ due to optimum ferromagnetism in FMTI is reported to be $\Delta \approx 28 - 50$ meV.[16,17] Moreover, departure from the $\sigma_{xy} = \pm \frac{e^2}{h} : \rho_{xx} = 0$ condition with increasing temperatures, while initially involving a complex phenomenology, consistently undergoes a transition to a $\sigma_{xx} \propto e^{-\frac{T_0}{T}}$ Arrhenius dependence[7,8,12,13,14,15] in the T~1 K range. How this could occur for states protected by a $\Delta > 28$ meV energy gap ($\frac{\Delta}{k_B} > 320$ K) is a conundrum[3,4] whose solution governs the long-term utility of QAH effects in these materials.

The surface states of three-dimensional TI exhibit a spectrum of delocalized states with Hamiltonian $H(\mathbf{k}) = \hbar v_F \mathbf{k} \cdot \boldsymbol{\sigma}$ yielding 2-dimension electron gas with a Dirac spectrum and spin-momentum locking.[18] In a FMTI with homogeneous magnetization $M_z$ parallel to the surface normal $\hat{z}$, an energy gap of magnitude $\Delta = J^*M_z/2\mu_B$ ($\equiv mv_F^2$) opens, where $m$ is the Dirac-mass, $\mu_B$ the Bohr magneton, $v_F$ the Fermi velocity, and $J^*$ the effective exchange energy. The resulting FMTI surface state spectrum is given by[19]

$$E_\pm(k) = E_D \pm \sqrt{(\hbar v_F)^2 k^2 + \Delta^2} \qquad (1)$$

where $E_D$ is the Dirac point of the ungapped bands measured relative to the surface-state $E_F$. Then, to achieve, detect and use the QAH effect and associated dissipationless transport, $E_F$ must satisfy $E_D - \Delta < E_F < E_D + \Delta$.

From a material science and technological applications point of view, the key challenges are therefore to create a long-range robust Dirac mass gap $\Delta$, in the SS of a TI with strong ferromagnetic order and whose bulk is truly insulating, and then to alter the surface electron density so as to place $E_F$ within that gap. A widely-explored strategy involves magnetic doping (e.g., with Cr or V atoms) of TIs (e.g., $Sb_2Te_3$) along with $E_F$ tuning with charge dopants (e.g., Bi). But such substitutional doping must necessarily create disorder[4] that is widely hypothesized[20,21,22] to be the limiting factor for observing QAH effect at higher temperatures. Amelioration strategies, such as remote magnetic doping[8] and co-doping[23] of the non-magnetic $(Bi_xSb_{1-x})_2Te_3$ have increased the QAHE observation temperatures slightly to almost 2 K, but this remains far smaller than what would be expected conventionally, given the large mass gap and Curie temperatures $T \gtrsim 100$ K. [9, 24] Thus, determining and understanding the combined atomic-scale effects of both the magnetic and charge dopant atoms on phenomenology of $E_D(\mathbf{r})$ and $\Delta(\mathbf{r})$ is now of critical interest.

To explore the electrostatic disorder induced in the SS of $Sb_2Te_3$ compounds, we employ LL spectroscopy[25,26,27,28,29,30] to simultaneously determine the Dirac energy $E_D$, the Dirac mass gap $\Delta$, and the Fermi velocities $v_F$ at nanoscale. In general, under an external magnetic field perpendicular to the sample surface, the SS becomes Landau quantized[28,29] with LL energies $E_n = E_D + \lambda\sqrt{2e\hbar B|n|v_F^2 + \Delta^2}$ for $n \neq 0$, where $B$ is the external magnetic field and $\Delta$ includes contributions from both the

exchange gap and the Zeeman gap. Here, $\lambda = 1$ for $n > 0$, and $\lambda = -1$ for $n < 0$. The 0th LL occurs at $E_0 = E_D \pm \Delta$, where the sign depends on the relative direction of the sample magnetization and the external magnetic field[28] – they being parallel in our studies. Hence, by measuring the set of LL energies $E_n(\mathbf{r})$ at each spatial location $\mathbf{r}$ and assuming LL quantization equations remain valid at nanoscale one can write

$$E_n(\mathbf{r}) = E_D(\mathbf{r}) + \lambda\sqrt{2e\hbar B|n|v_F^2(\mathbf{r}) + \Delta(\mathbf{r})^2} \text{ for } n \neq 0 \qquad (2)$$

so that

$$E_D(\mathbf{r}) = \frac{E_0^2 + E_{-2}^2 - 2E_{-1}^2}{2(E_0 + E_{-2} - 2E_{-1})} \qquad (3)$$

and

$$\Delta(\mathbf{r}) = E_0(\mathbf{r}) - E_D(\mathbf{r}) \qquad (4)$$

Here, the Fermi velocities of the top and bottom Dirac cones are not assumed identical to account for any electron-hole asymmetry. In the special case of electron-hole symmetry, Eqn. (3) is simplified as $E_D(\mathbf{r}) = (E_{-1} + E_1)/2$. Furthermore, a linear dispersion of the topological surface state is assumed because the relevant LLs are within $\pm 70$ meV of the Dirac point, where the dispersion has been shown to be linear.[17,31] Our strategy is then to compare and contrast the electronic structure of the SS in $(Bi_{0.1}Sb_{0.9})_2Te_3$ and $Cr_{0.08}(Bi_{0.1}Sb_{0.9})_{1.92}Te_3$ single crystals, using nanoscale LL spectroscopy[25,26,27,28,29] techniques in a spectroscopic imaging scanning tunneling microscope (SISTM)[32] operated at $T \lesssim 300\ mK$ and in magnetic fields up to 9 T.

The $(Bi_{0.1}Sb_{0.9})_2Te_3$ crystals consist of stacked quintuple layers with 10% of the Sb sites occupied by Bi substitutional dopant atoms in a random manner as shown schematically in Fig. 1a. Once cleaved, the crystal is terminated by Te atoms arranged in a triangular lattice with nearest-neighbor atoms separated by 4.3 Å (Fig. 1b). This lattice forms the main contrast in atomic-resolution topographic images shown in Fig. 1b. The subsurface Bi dopant atoms lead to heterogeneous topography with spatially varying bright spots (inset Fig. 1b). Electronic structure visualization is then carried out by measuring the differential tunneling conductance $dI/dV(\mathbf{r}, E = eV) \equiv g(\mathbf{r}, E)$ as a function of both location $\mathbf{r}$ and electron energy $E$. At each $\mathbf{r}$, the local density of electronic states $N(\mathbf{r}, E)$ is conventionally determined from $N(\mathbf{r}, E) \propto g(\mathbf{r}, E)$. For these materials, the $g(\mathbf{r}, E)$ data are dominated by bulk bands at high energy and by the topological SS for $0 \lesssim E \lesssim 400$ meV. Thus one can evaluate the local effects of Bi dopant disorder on the bulk bands by measuring spatial variants in $N(\mathbf{r}, E)$ at high energies. Figure 1c shows a pair of simultaneous $g(\mathbf{r}, E)$ images acquired at $E = -0.4$ eV and $E = 0.8$ eV (at zero magnetic field). Underpinning these phenomena are the high-energy shifts in $g(\mathbf{r}, E)$ discussed in Figure 1d. Here the blue and red $g(E)$ curves are averages from the regions indicated by the blue and red circles in Fig. 1c. These two $g(E)$ curves are laterally displaced while their functional form remains virtually unchanged, indicating that the images in Fig. 1d represent rigid bulk band shifts between different regions of the sample[33].

Next the magnetic field is turned on and the $g(\mathbf{r}, E)$ mapping of LL is carried out. Figure 2a shows the spatially averaged $g(E)$ spectrum of $(Bi_{0.1}Sb_{0.9})_2Te_3$ under an 8.5 T magnetic field after a background subtraction (raw spectrum shown in Fig. S1). The peaks due to LL quantization of the SS are clearly resolved as indicated. Since the Zeeman energy is negligible at this field, $\Delta$ is taken as zero for this non-magnetic TI. Therefore, spatial fluctuations of $E_D(\mathbf{r})$ are derived from $E_0(\mathbf{r})$ in Eqn. (4) with their statistics shown in Fig. 2b. On average, the LL-derived $E_D$ occurs at 143.4 meV with a standard deviation $\sigma \cong 3.3$ meV (FWHM is $2.35\sigma \approx 7.8$ meV), indicative of a significant electrostatic spatial variations in $E_D$ of the SS. Such $E_D$ disorder is absent in pure $Sb_2Te_3$ crystals without the Bi dopant atoms.[28] For $(Bi_{0.1}Sb_{0.9})_2Te_3$ samples, by contrast, the Dirac energy disorder is visualized by plotting measured $E_D(\mathbf{r})$ as shown in Fig. 2c. Moreover, Figure 2d shows the measured and calculated rigid band shift $\Gamma(\mathbf{r})$ in the region of the white square in Figure 2c (see details of calculation in Fig. S2). The features correspond well, suggesting that the Dirac point and bulk bands locally shift in the same direction. Such observation is in agreement with macroscopic band structure characterization of $(Bi_xSb_{1-x})_2Te_3$ with different nominal Bi concentrations using angle-resolved photoemission spectroscopy.[31] By using measured $E_n(\mathbf{r})$ (Fig. 2a), the Fermi velocity of the top and bottom Dirac cones can also be determined as a function of position (see Fig. S3). The Fermi velocities $v_{FT}(\mathbf{r})$ and $v_{FB}(r)$ show spatial fluctuations and clear anticorrelation with each other. They also anticorrelate and correlate with $E_D(r)$ respectively or $\delta E_D \delta v_{FT} < 0$ and $\delta E_D \delta v_{FB} > 0$, where $\delta$ denotes variations. Specifically, an upwards $E_D$ shift (i.e., $\delta E_D > 0$) is found associated with $\delta v_{FB}(r) > 0$ and $\delta v_{FT}(r) < 0$, and vice versa. In addition to the averaging effect from larger radius of higher LL wavefunctions, this form of SS band structure variations also explains the observed narrower distribution of higher LLs (i.e., larger $n$) compared to that of the 0$^{th}$ LL considering that $\frac{\delta E_n}{\delta E_0} = 1 + \sqrt{2e\hbar B n}\frac{\delta v_{FT}}{\delta E_D}$ for $n > 0$ and $\frac{\delta E_n}{\delta E_0} = 1 - \sqrt{2e\hbar B |n|}\frac{\delta v_{FB}}{\delta E_D}$ for $n < 0$ (see Fig. S4).

Having characterized the electrostatic disorder generated by heterogeneity of the Bi dopant atoms in (Bi$_{0.1}$Sb$_{0.9}$)$_2$Te$_3$, we next turn to explore the combined effect of electrostatic and magnetic dopants in Cr$_{0.08}$(Bi$_{0.1}$Sb$_{0.9}$)$_{1.92}$Te$_3$. The Cr dopant atoms appear as additional well-defined dark triangles in topographic images (Fig. 3a), in agreement with previous reports.[17,24] Here, we determine both $E_D(r)$ and $\Delta(r)$ by measuring LL energies as shown in the background-subtracted spectrum in Fig. 3b (raw spectrum shown in Fig. S5). Compared to asymmetric LLs caused by electron-hole asymmetry (e.g., different Fermi velocities) in (Bi$_x$Sb$_{1-x}$)$_2$Te$_3$ (Fig. 2 and Fig. S3) and twisted graphene layers,[34] the asymmetric LLs in the case of a Dirac-mass gap opening is manifested as an energy shift of the zeroth LL by the amount of the gap. Figure 3c shows measured $E_D(r)$ and demonstrates it to be statistically highly consistent with $E_D(r)$ in (Bi$_{0.1}$Sb$_{0.9}$)$_2$Te$_3$. Then the Dirac mass gap $\Delta(r)$ is determined using Eqn. (4) with the typical result shown in Figure 3d. The spatial structure of $E_D(r)$ and $\Delta(r)$ in the same FOV are virtually uncorrelated with maximum cross correlation coefficient of $\chi(r = 0) = -0.23$. This is expected given that Cr dopants resulting in $\Delta(r)$ disorder contributes little to $E_D(r)$ disorder due to Bi, and vice versa.[28]

The simultaneously measured histograms of $E_D(r)$ and $\Delta(r)$ are given in Figures 4a,b. The average of $E_D(r)$ in Cr$_{0.08}$(Bi$_{0.1}$Sb$_{0.9}$)$_{1.92}$Te$_3$ is 171.2 meV with standard deviation $\sigma \cong 2.9$ meV (FWHM is $2.35\sigma = 6.83$ meV), close to that of $(Bi_{0.1}Sb_{0.9})_2Te_3$. The Dirac-mass gap Δ centers near 14 meV with a minimum value of 4.7 meV. Compared to preliminary Dirac-mass gap extraction from zero-field d$I$/d$V$ spectra,[17] which revealed a consistent phenomenology of Δ disorder, LL spectroscopy is a more fundamental technique that is expected to yield quantitatively more accurate measures of Δ. Although the standard deviation of $\Delta(\mathbf{r})$, $\sigma \cong 1.7$ meV, is slightly smaller than that of $E_D(\mathbf{r})$, the more peaked shape of Δ than $E_D$ results in a similar distribution width of ~ 10 meV. This suggests a similar magnitude in the electrostatic and magnetic disorder in the Dirac SS in these doped $Sb_2Te_3$ FMTI materials. Indeed, by plotting the top and bottom band edges of the gapped surface state ($E_D(\mathbf{r}) + \Delta(\mathbf{r})$ and $E_D(\mathbf{r}) - \Delta(\mathbf{r})$, respectively) as shown in Figure 4 c,d and their energy distributions as shown in Figure 4e, the minimum distance between the two energy-surfaces is only ~3.8 meV, much smaller than the average gap in earlier reports.[16,17] This energy scale, however, is not what is critical in thermally activated transport. Instead, because the SS Fermi level $E_F$ is inserted into this gap, it is the spatial arrangement of the minimum energy separation of positive or negative gap edge to $E_D$. This corresponds to a minimum effective Dirac-mass gap of less than 1.9 meV within this 200 nm field of view, and with spatial variation caused not merely by the magnetic dopant atoms but also by the charge dopant atoms. One can measure this minimum excitation energy difference between the chemical potential and the local gap edge, either above or below $E_F$, as a function of $\mathbf{r}$: $\delta(\mathbf{r}) = \min(\Delta(\mathbf{r}) + E_D(\mathbf{r}) - E_F; \Delta(\mathbf{r}) - E_D(\mathbf{r}) + E_F)$, where $E_F$ is assumed at the optimum position maximizing $\min(\delta(\mathbf{r}))$. Since $\Delta(\mathbf{r})$ and $E_D(\mathbf{r})$ are not correlated, the $\delta(\mathbf{r})$ shown in Figure 4f represents the previously unknown spatial arrangement of energy barriers to activated electric transport in $Cr_{0.08}(Bi_{0.1}Sb_{0.9})_{1.92}Te_3$ samples that sustain the QAH effect.

The statistical distribution of our measured $\delta(\mathbf{r})$ is shown in Figure 4g and well fitted with a skewed normal distribution function, which is confirmed with our numerical simulations (Fig. S6):

$$f(\delta) = \frac{2}{\sigma\sqrt{2\pi}} e^{-\frac{1}{2}(\frac{\delta-\mu}{\sigma})^2} \int_{-\infty}^{\alpha(\frac{\delta-\mu}{\sigma})} \frac{1}{\sqrt{2\pi}} e^{-\frac{1}{2}x^2} dx \qquad (5)$$

where $\mu = 13.54$ meV, $\sigma = 3.65$ meV, $\alpha = -1.90$. Such distribution allows us to consider $\delta(r)$ beyond the field of view limited by our SISTM. For example, while the probability density falls off to zero quickly for $\delta \gg u$, the long tail of $\delta \ll u$ suggests non-negligible probability of $\delta(r)$ approaching zero mass gap. A more detailed analysis of this low-energy tail in Fig. 5a reveals $\delta(r) < 100$ μeV with a probability $P \sim 2 \times 10^{-4}$. This implies that, statistically, regions where the energy gap falls below ~100 μeV are separated by $d \cong \sqrt{\frac{\xi^2}{P}} = 0.85$ μm in this material, where $\xi = 12.1$ nm is the coherence length of $\delta(r)$ determined from Fig.4f representing the characteristic length over which $\delta(r)$ varies. Figure 5b shows a schematic of these effects and it is in this physical context that one must consider the departure of QAH effect from the $\sigma_{xy} = \pm \frac{e^2}{h} : \rho_{xx} = 0$ condition, with increasing temperatures. With an average mass gap above 10 meV, thermally activated band conduction ($\sigma_{xx} \propto e^{-\frac{\Delta}{k_B T}}$) should be tremendously suppressed at 1 K. On the other hand, thermally activated hopping transport through nearest-neighbor of localized states would result in a similar exponential form of $\sigma_{xx} \propto e^{-\frac{T_0}{T}}$, but require a far lower activation energy ($k_B T_0 \ll \Delta$), where $k_B T_0$ is the activation energy on the order of the energy separation of neighboring localized states. Indeed, transport measurements of QAH materials at temperatures around 1 K consistently revealed an activation energy ranging on the order of 100 μeV.[12,13,14,15] In lightly doped semiconductors, nearest-neighbor hopping is preferred when the inter-donor distance is much larger than the effective Bohr radius of the dopants[35]. Given the reported localization length of hundreds of nanometers[13] in $Cr_x(Bi_{0.1}Sb_{0.9})_{2-x}Te_3$, it is therefore plausible that charges hop between the localized states of the surface state band extrema, which our data indicate are separated by ~0.8 μm with activation energies near 100 μV (Fig. 5b). At even lower temperatures $k_B T \ll 100$ μeV, this nearest-neighbor hopping should transition into variable-range hopping with a different functional form of $\sigma_{xx} \propto e^{-(\frac{T_M}{T})^{1/3}}$, which has also been observed experimentally in $Cr_{0.08}(Bi_{0.1}Sb_{0.9})_{1.92}Te_3$ at temperatures below 200 mK.[4,12] The dissipative nature of such transport channels resulting from electrostatic and magnetic heterogeneity then limits the observation of QAH effect to temperatures well below $\langle\Delta\rangle/k_B$ and the Curie temperature.

In summary, by utilizing LL spectroscopy at sub-Kelvin temperatures, we have examined separately and at atomic scale, the electrostatic and magnetic disorder due to individual dopant atoms in the canonical QAH effect materials $(Bi_{0.1}Sb_{0.9})_2Te_3$ and $Cr_{0.08}(Bi_{0.1}Sb_{0.9})_{1.92}Te_3$. This reveals that electrostatic disorder randomizes the Dirac energy while the magnetic disorder independently randomizes the Dirac mass gap (Fig. 3). Together these two phenomena conspire to drastically suppress the minimum Dirac mass gap by two orders of magnitude from the mean, in regions separated by less than a micron (Figs 4, 5), plausibly rendering thermally activated transport possible at temperatures around 1 K. Our study thus provides a more complete picture of the material characteristics underpinning the very low temperatures where the QAH effect can be observed in these materials. The key implication is that, for practical applications of QAH insulators at higher temperatures, homogenization of both the SS band structure and of the Dirac mass gap is required, for example, by engineering intrinsic magnetic TIs.[36] Meanwhile, efforts towards engineering FMTIs with larger Dirac-mass gaps would effectively mitigate the influence of electrostatic disorders.

**Materials and Methods:**

Single crystals of $Cr_{0.08}(Bi_{0.1}Sb_{0.9})_{1.92}Te_3$ and $(Bi_{0.1}Sb_{0.9})_2Te_3$ were synthesized using a floating-zone method, where high purity Bi, Sb, Te, Cr (for $Cr_{0.08}(Bi_{0.1}Sb_{0.9})_{1.92}Te_3$) were sealed in double-walled quartz ampoules under vacuum conditions and melted at 900°C. The premelt ingot rods in quartz tubes were then mounted in a floating-zone furnace and premelted at a rate of 200 mm/h. Then the crystals were grown at a rate of 1 mm/h in a 1 bar Ar atmosphere.

STM experiments were performed in a home-built system in an ultra-high vacuum environment. Bulk crystals were mechanically cleaved *in situ* at temperatures below 10 K in vacuum and immediately transferred to STM head for characterization using tungsten tips. The tips were prepared by field-emission on Au substrates until a flat density-of-state near the Fermi level was obtained. Spectroscopic characterizations of the samples were performed at a base temperature of 300 mK with a single-shot $^3$He cryostat. A lock-in amplifier (SR830) was used to obtain differential conductance data with a modulation frequency of ~850 Hz and modulation amplitude of 80 μV.

**Acknowledgements**: We acknowledge and thank A. V. Balatsky, V. Madhavan, and S. Simon for very helpful discussions and communications. Y.X.C., A.K., R.S., G.G., and K.F. acknowledge support from the US Department of Energy, Office of Basic Energy Sciences, under contract number DEAC02-98CH10886. X.L. acknowledges support from the Kavli Institute at Cornell through the KIC Postdoctoral Fellowship and from the Keck Foundation. J.C.S.D and P.O.S. acknowledge support from the Gordon and Betty Moore Foundation's EPiQS Initiative through Grant GBMF4544. J.C.S.D acknowledges support from Science Foundation Ireland under Award SFI 17/RP/5445, and from the European Research Council (ERC) under Award DLV-788932.

**Author Contributions**: Y.X.C and X.L contributed equally to this work.

**Notes**: The authors declare no conflict of interest.

**Supporting Information**:
This material is available free of charge via the internet at http://pubs.acs.org
1. Determination of Fermi velocities and the Dirac point
2. Discussion on the narrower distribution of higher LLs compared to that of the 0$^{th}$ LL
Figure S1. Raw LL spectrum of $(Bi_{0.1}Sb_{0.9})_2Te_3$ and its uniform background used for subtraction.
Figure S2. Bulk band shift as a function of position in $(Bi_{0.1}Sb_{0.9})_2Te_3$.
Figure S3. Spatial maps of Fermi velocities.

Figure S4. Evolution of LLs in $(Bi_{0.1}Sb_{0.9})_2Te_3$.

Figure S5. Raw LL spectrum of $Cr_{0.08}(Bi_{0.1}Sb_{0.9})_{1.92}Te_3$ and its uniform background used for subtraction.

Figure S6. Skewed normal distribution of simulated $\delta(r)$.

Figure S7. Comparison of $E_D(r)$ extracted using different LLs and zero magnetic field data.

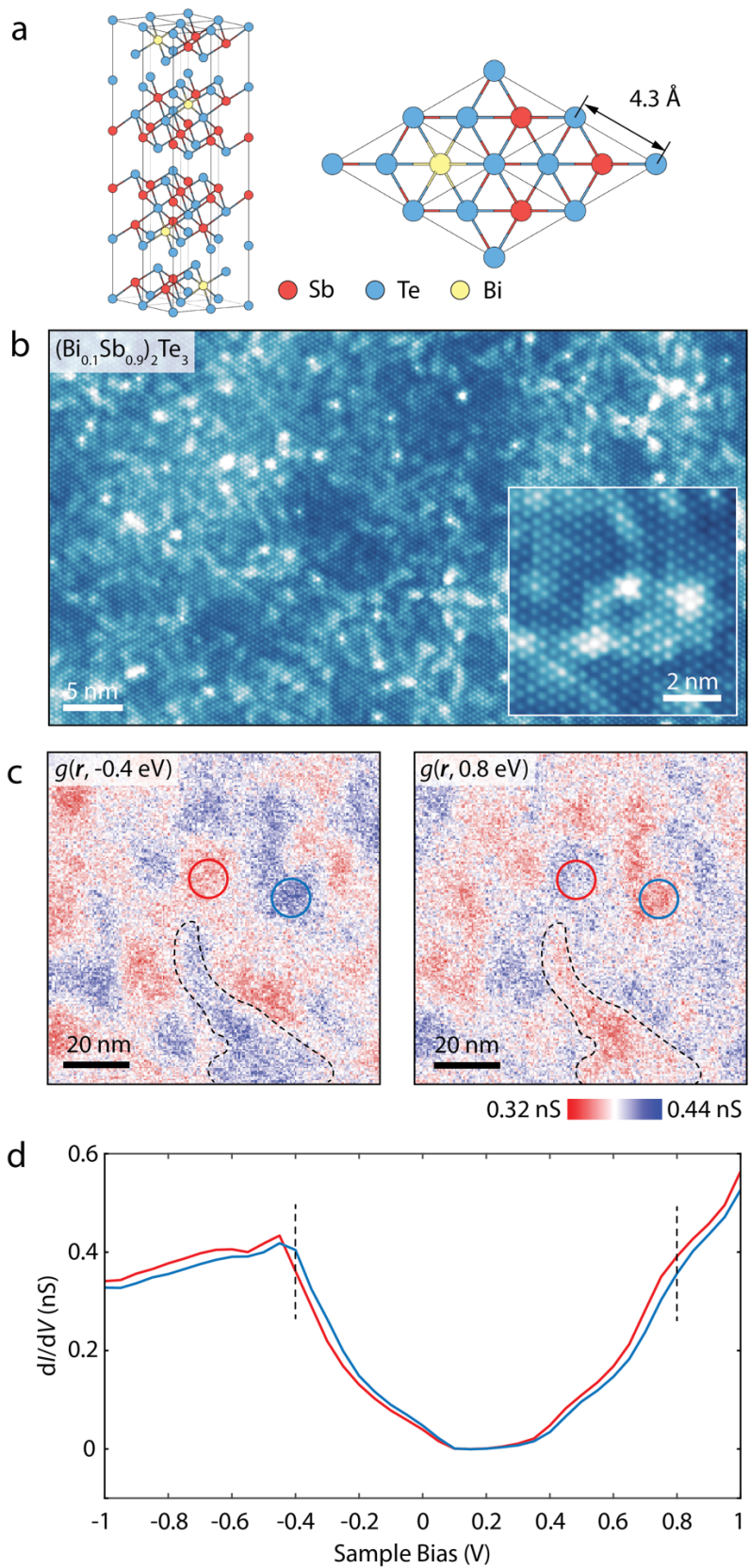

**Figure 1.** Spectroscopic measurements of $(Bi_{0.1}Sb_{0.9})_2Te_3$ single crystals.

(a) Schematics of $(Bi_{0.1}Sb_{0.9})_2Te_3$ crystal structure in 3D (left) and its in-plane lattice structure as viewed from above (right).

(b) STM topographic images of Te termination layer of $(Bi_{0.1}Sb_{0.9})_2Te_3$ with inset being a zoomed-in image.

(c) Differential conductance $g(r,E)$ maps of the same region at -0.4 eV and 0.8 eV.

(d) Spatially averaged differential conductance $g(E)$ spectra of $(Bi_{0.1}Sb_{0.9})_2Te_3$ acquired in a zero magnetic field. The red and blue curves are averaged spectra taken from the regions marked by the red and blue circles in C, respectively. The dashed curves correspond to the same region in two inset images.

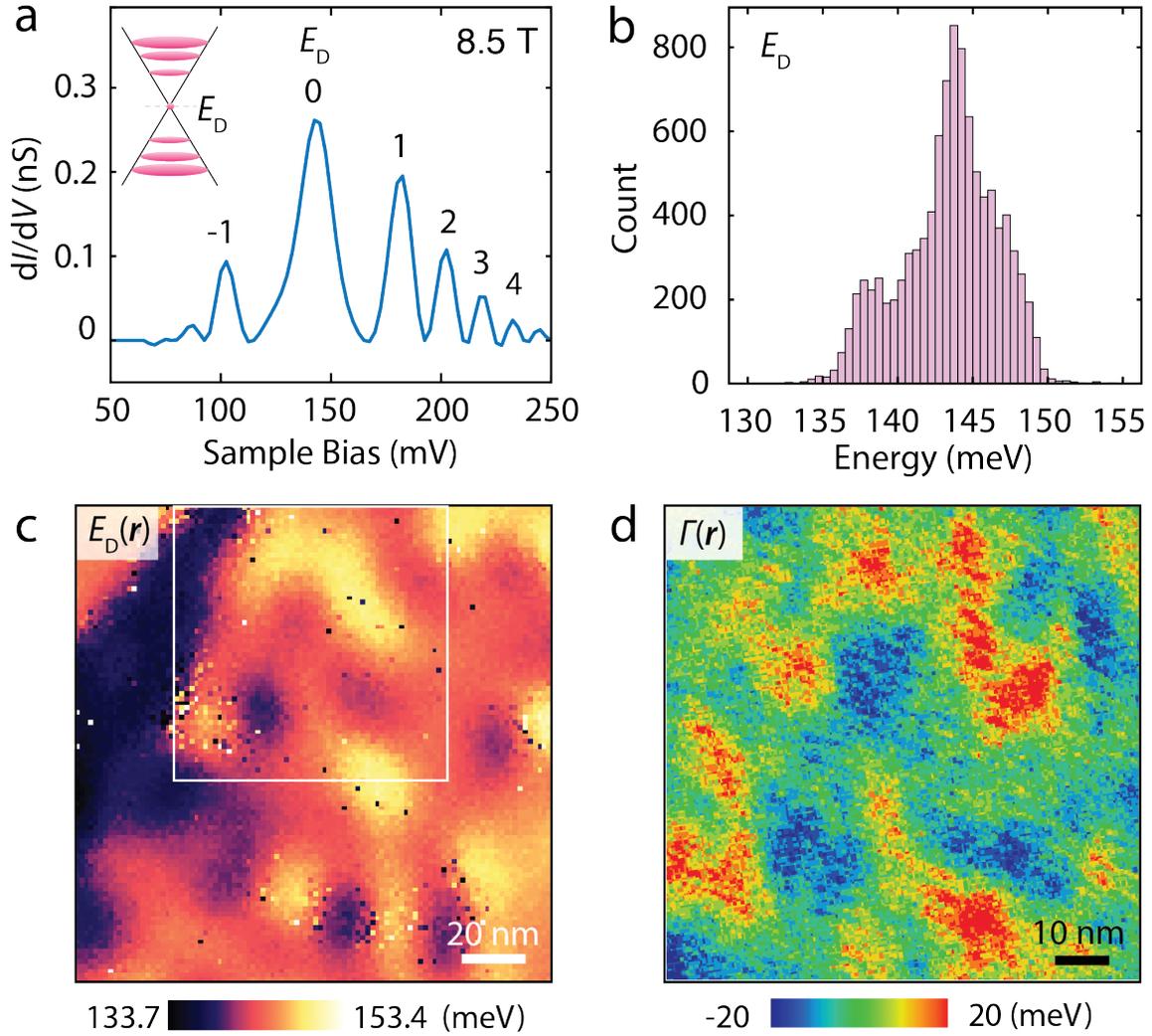

**Figure 2.** Correlated disorders in the Dirac surface states of $(Bi_{0.1}Sb_{0.9})_2Te_3$.

(a) Averaged differential conductance spectrum of $(Bi_{0.1}Sb_{0.9})_2Te_3$ taken in an 8.5 T magnetic field showing well-defined LLs. The numbers correspond to LL indices $n$.

(b) Typical histogram of measured $E_D$ for $(Bi_{0.1}Sb_{0.9})_2Te_3$.

(c) Typical $E_D(\boldsymbol{r})$ image showing spatial arrangements of disorder in the Dirac energy of the SS of $(Bi_{0.1}Sb_{0.9})_2Te_3$.

(d) Typical rigid band shift $\Gamma(\boldsymbol{r})$ image showing spatial arrangements of disorder in the bulk energy bands of $(Bi_{0.1}Sb_{0.9})_2Te_3$. $\Gamma(\boldsymbol{r})$ is extracted from differential conductance spectra.

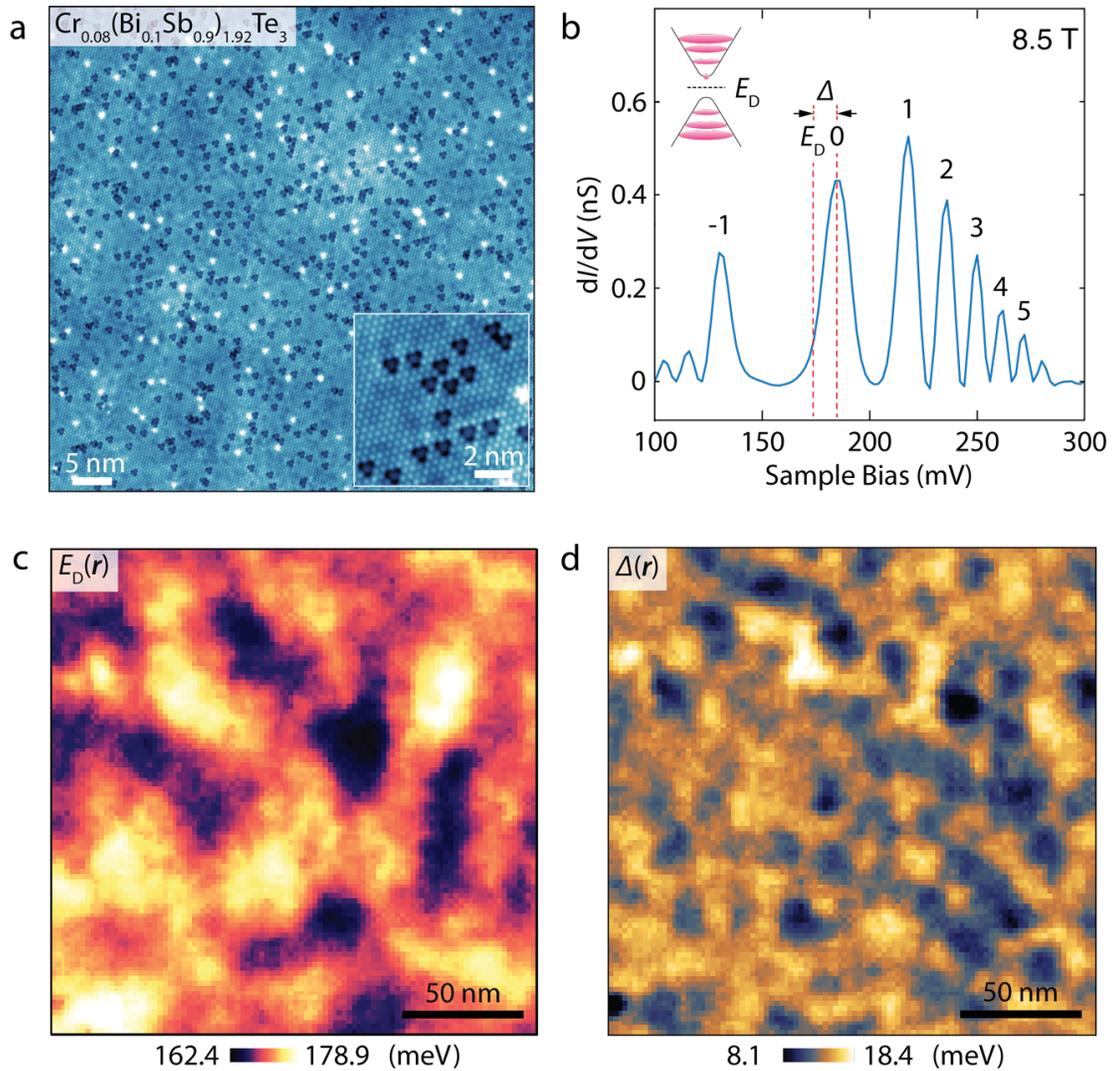

**Figure 3** Topographic and spectroscopic characterizations of $Cr_{0.08}(Bi_{0.1}Sb_{0.9})_{1.92}Te_3$.

(a) STM topographic images of $Cr_{0.08}(Bi_{0.1}Sb_{0.9})_{1.92}Te_3$ with the inset showing a zoomed-in image.

(b) Averaged differential conductance spectrum taken in an 8.5 T magnetic field showing well-defined LLs of $Cr_{0.08}(Bi_{0.1}Sb_{0.9})_{1.92}Te_3$.

(c) Typical $E_D(r)$ image showing spatial arrangements of disorder in the Dirac energy of the SS of $Cr_{0.08}(Bi_{0.1}Sb_{0.9})_{1.92}Te_3$..

(d) Typical $\Delta(r)$ image showing spatial arrangements of disorder in the Dirac mass gap of $Cr_{0.08}(Bi_{0.1}Sb_{0.9})_{1.92}Te_3$.

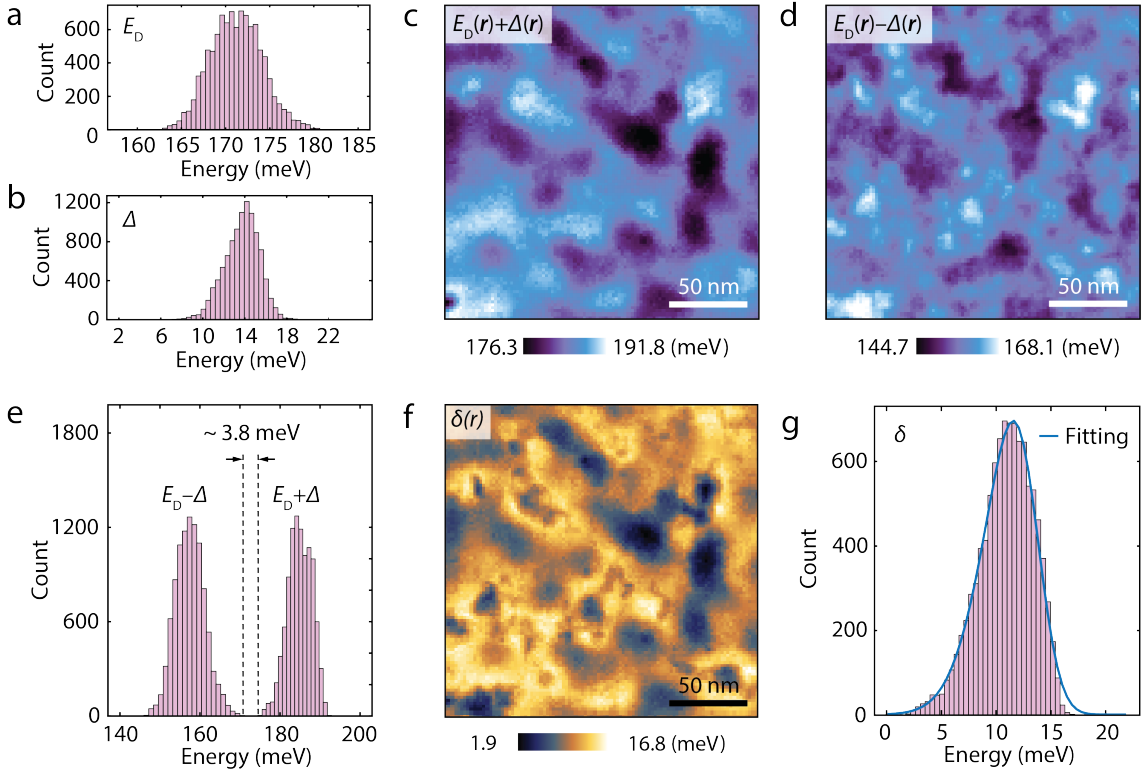

**Figure 4** How charge and magnetic dopant atoms conspire to degrade SS band structure of $Cr_{0.08}(Bi_{0.1}Sb_{0.9})_{1.92}Te_3$.

(a, b) Measured histograms of $E_D$ and $\Delta$

(c, d) Measured images of $E_D(r) + \Delta(r)$ and $E_D(r) - \Delta(r)$

(e) Measured histograms of $E_D(r) + \Delta(r)$ and $E_D(r) - \Delta(r)$ from C, D.

(f) Measured image of $\delta(r) = \min(\Delta(r) + E_D(r) - E_F;\ \Delta(r) - E_D(r) + E_F)$. Comparing $E_D(r)$ in Figure 3e and $\delta(r)$ in Figure 4f, both the local minima and local maxima in $E_D(r)$ correspond to local minima in $\delta(r)$. This suggests any significant local variations in $E_D$ would results in the reduction of local effective energy barrier for activated transport.

(g) Measured histogram of $\delta(r)$ and fitting using a skewed normal distribution.

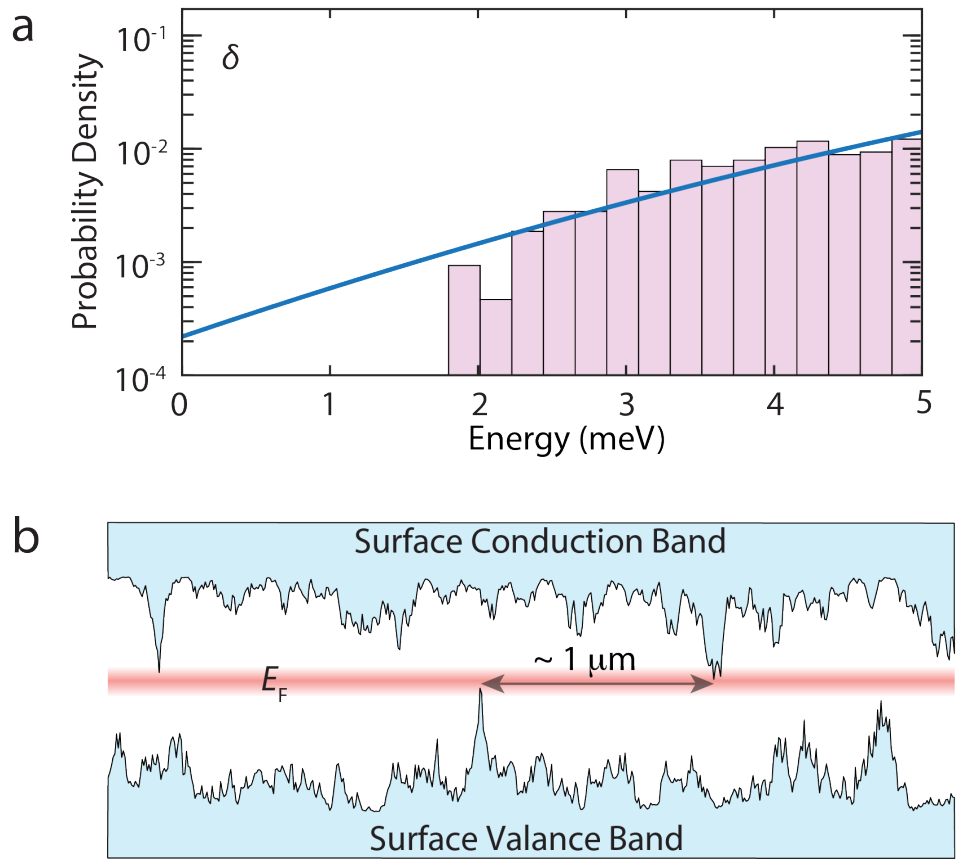

**Figure 5** Statistics for $\delta(r)$ distribution in $Cr_{0.08}(Bi_{0.1}Sb_{0.9})_{1.92}Te_3$.

(a) Low energy portion of the histogram for the probability density of $\delta(r)$ from Figure 4g in log scale.

(b) Schematics of surface valence and conduction band heterogeneity.

**TOC figure**

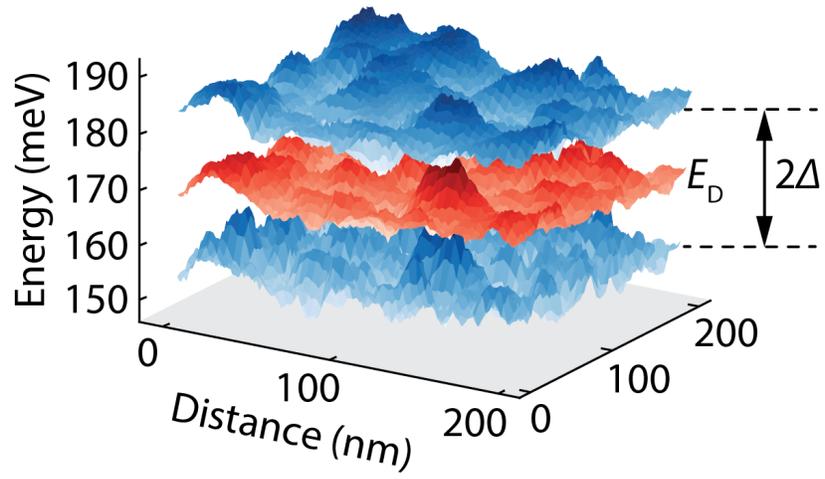

**References**

1   Haldane, D. Model for a Quantum Hall Effect without Landau Levels: Condensed-Matter Realization of the "Parity Anomaly", *Phys. Rev. Lett.* **1988**, *61*, 2015.

2   Ogushi, K., Murakami, S. and Nagaosa, N. Spin Anisotropy and Quantum Hall Effect in the Kagomé Lattice: Chiral Spin State Based on a Ferromagnet. *Phys. Rev. B* **2000**, *62*, R6065.

3   Tokura, Y., Yasuda, K., Tsukazaki, A. Magnetic Topological Insulators. *Nat. Rev. Phys.* **2019**, *1*, 126–143.

4   Chang, C.-Z. *et al.*, Experimental Observation of the Quantum Anomalous Hall Effect in a Magnetic Topological Insulator. *Science* **2013**, *340*, 167–170.

5   Checkelsky, J.G. *et al.*, Trajectory of the Anomalous Hall Effect Towards the Quantized State in a Ferromagnetic Topological Insulator. *Nat. Phys.* **2014**, *10*, 731–736.

6   Kou, X. *et al.*, Scale-Invariant Quantum Anomalous Hall Effect in Magnetic Topological Insulators Beyond the Two-Dimensional Limit. *Phys. Rev. Lett.* **2014**, *113*, 137201.

7   Bestwick, A. J. *et al.*, Precise Quantization of Anomalous Hall Effect Near Zzero Magnetic Field. *Phys. Rev. Lett.* **2015**, *114*, 187201.

8   Mogi, M. *et al.*, Magnetic Modulation Doping in Topological Insulators Toward Higher-Temperature Quantum Anomalous Hall Effect. *Appl. Phys. Lett.* **2015**, *107*, 182401.

9   Chang, C.-Z. *et al.*, High-Precision Realization of Robust Quantum Anomalous Hall State in a Hard Ferromagnetic Topological Insulator. *Nat. Mater*, **2015**, *14*, 473.

10  Feng, Y. *et al.* Observation of the Zero Hall Plateau in a Quantum Anomalous Hall Insulator. *Phys Rev Lett* **2015**, *115*, 126801.

11  Yasuda, K. *et al.,* Quantized Chiral Edge Conduction on Domain Walls of a Magnetic Topological Insulator. *Science*, **2017**, *358*, 1311-1314.

12  Kawamura, M. *et al.*, Current-Driven Instability of the Quantum Anomalous Hall Effect in Ferromagnetic Topological Insulators. *Phys. Rev. Lett.* **2017**, *119*, 016803.

13 Fox, E. J. *et al.,* Part-Per-Million Quantization and Current-Induced Breakdown of the Quantum Anomalous Hall Effect. *Phys. Rev. B*, **2018**, *98*, 075145.

14 Ou, Y. *et al.*, Enhancing the Quantum Anomalous Hall Effect by Magnetic Codoping in a Topological Insulator. *Adv. Mater.* **2018**, *30*, 1703062.

15 Rosen, I.T. *et al.* Chiral Transport along Magnetic Domain Walls in the Quantum Anomalous Hall Effect. *npj Quantum Materials* **2017**, *2*, 69.

16 Chen, Y. L. *et al.* Massive Dirac Fermion on the Surface of a Magnetically Doped Topological Insulator. *Science* **2010**, *329*, 659-662.

17 Lee, I. *et al.*, Imaging Dirac-Mass Disorder from Magnetic Dopant Atoms in the Ferromagnetic Topological Insulator $Cr_x(Bi_{0.1}Sb_{0.9})_{2-x}Te_3$. *Proc. Natl. Acad. Sci. U.S.A* **2015**, *112*, 1316-1321.

18 Hasan, M. Z. and Kane, C. L. Colloquium: Topological insulators, *Rev. Mod. Phys.* **2010**, *82*, 3045.

19 Luo, W. and Qi, X.-L. Massive Dirac Surface States in Topological Insulator/Magnetic Insulator Heterostructures, *Phys. Rev. B* **2013**, *87*, 085431.

20 Feng, X. *et al.*, Thickness Dependence of the Quantum Anomalous Hall Effect in Magnetic Topological Insulator Films, *Adv. Mater.* **2016**, *28*, 6386-6390.

21 Yue Z. and Raikh M. E., Smearing of the Quantum Anomalous Hall Effect due to Statistical Fluctuations of Magnetic Dopants. *Phys. Rev. B* **2016**, *94*, 155313.

22 He, K., Wang, Y., and Xue, Q.-K. Topological Materials: Quantum Anomalous Hall System. *Annu. Rev. Condens. Matter. Phys*. **2018**, *9*, 329-344.

23 Y. Ou, *et al.*, Enhancing the Quantum Anomalous Hall Effect by Magnetic Codoping in a Topological Insulator. *Adv. Mater.* **30**, 1703062 (2018).

24 Chang, C.-Z. *et al.*, Thin Films of Magnetically Doped Topological Insulator with Carrier-Independent Long-Range Ferromagnetic Order. *Adv. Mater.* **2013**, *25*, 1065–1070.

25 Cheng, P. *et al.*, Landau Quantization of Topological Surface States in $Bi_2Se_3$, *Phys. Rev. Lett.* **2010**, *105*, 076801.

26 Okada, Y. *et al.*, Observation of Dirac Node Formation and Mass Acquisition in a Topological Crystalline Insulator, *Science* **2013**, *341*, 1496-1499.

27 Hanaguri, T., Igarashi, K., Kawamura, M., Takagi, H., Sasagawa, T. Momentum-Resolved Landau-Level Spectroscopy of Dirac Surface State in $Bi_2Se_3$, *Phys. Rev. B* **2010**, *82*, 081305(R).

28 Jiang, Y. *et al.*, Mass Acquisition of Dirac Fermions in Magnetically Doped Topological Insulator $Sb_2Te_3$ Films, *Phys. Rev. B* **2015**, *92*, 195418.

29 Sessi, P. *et al.*, Dual Nature of Magnetic Dopants and Competing Trends in Topological Insulators, *Nat. Commun*. **2016**, *7*, 12027.

30 Li, G. and Andrei, E. Observation of Landau levels of Dirac fermions in graphite. *Nat. Phys*. **2007**, *3*, 623–627.

31 Zhang, J. *et al.*, Band Structure Engineering in $(Bi_{1-x}Sb_x)_2Te_3$ Ternary Topological Insulators. *Nat. Commun.* **2011**, *2*, 574.

32 Fujita, K. *et al*. Spectroscopic Imaging STM: Atomic-Scale Visualization of Electronic Structure and Symmetry in Underdoped Cuprates. *Strongly correlated systems: Experimental techniques*, eds Avella A, Mancini F (Springer, Heidelberg), pp 73-110.

33 Beidenkopf, H. *et al.*, Spatial Fluctuations of Helical Dirac Fermions on the Surface of Topological Insulators. *Nat. Phys.* **2011**, *7*, 939–943.

34 Luican, A. *et al.*, Single-Layer Behavior and its Breakdown in Twisted Graphene Layers. *Phys. Rev. Lett*. **2011**, *106*, 126802.

35 Y. Natsume, Y., and H. Sakata. Zinc Oxide Films Prepared by Sol-Gel Spin-Coating. *Thin Solid Films* **2000**, *372*, 30-36.

36 Deng, Y. *et al.*, Quantum Anomalous Hall Effect in Intrinsic Magnetic Topological Insulator $MnBi_2Te_4$. *Science* **2020**, *367*, 895-900.